# Applicability of Educational Data Mining in Afghanistan: Opportunities and Challenges

Abdul Rahman, Sherzad

Lecturer at Computer Science Faculty of Herat University, Afghanistan

Ph.D. Student at Technische Universität Berlin, Germany

absherzad@gmail.com || absherzad@mailbox.tu-berlin.de

**ABSTRACT**

The author's own experience as a student and later as a lecturer in Afghanistan has shown that the methods used in the educational system are not only flawed, but also do not provide the minimum guidance to students to select proper course of study before they enter the national university entrance (Kankor) exam. Thus, it often results in high attrition rates and poor performance in higher education.

Based on the studies done in other countries, and by the author of this paper through online questionnaires distributed to university students in Afghanistan – it was found that proper procedures and specialized studies in high schools can help students in selecting their major field of study more systematically.

Additionally, it has come to be known that there are large amounts of data available for mining purposes, but the methods that the Ministry of Education and Ministry of Higher Education use to store and produce their data, only enable them to achieve simple facts and figures. Furthermore, from the results it can be concluded that there are potential opportunities for educational data mining application in the domain of Afghanistan's education systems. Finally, this study will provide the readers with approaches for using Educational Data Mining to improve the educational business processes. For instance, predict proper field of study for high school graduates, or, identify first year university students who are at high risk of attrition.

**Keywords**

Educational data mining; data mining; major study prediction; placement class; computer uses in education; value of information; Kankor, University Entrance Exam, Afghanistan education systems.

## 1. INTRODUCTION

Recent advances in big data, data mining, and educational data mining (EDM) enable educational institutions to transform their data into valuable information to find the hidden patterns that yields clear results and specific targets to improve educational settings.

The efforts of educational institutions in Afghanistan are to generate (only) facts and figures (e.g., total number of students and teachers based on gender, geographic location, schools and universities, and some other conditions). However, it turned out that these simple facts and figures do not help educational institutions to improve the educational settings. For example, to predict proper major fields of study for high school graduates participating in the Kankor exam, to identify first year university students who are at high attrition risk, or to recommend right courses for the students to enroll.

To find out about participants' familiarity and factors choosing proper major studies in the Kankor exam, attrition rates and poor quality of education, and whether offering specialized studies at high school is a good decision or not; online questionnaires (open-ended and close-ended questions) were prepared and distributed through social networking platforms to Afghanistan public and private university students and alumni. Results of the questionnaires indicated that lack of familiarity with the Kankor exam and its processes, inappropriate selection of major fields of study as well as lack of offering specialized studies at school are the major reasons that most of the students are at risk of dropping out or poor performance in their higher education studies.

The main goal and objective of this paper is to study the opportunities and challenges of EDM applicability in Afghanistan education context to help educational institutions to better prepare students for their studies in schools and universities. This paper also tries to offer answers to the following questions:

1. What are the potential challenges to implement EDM in the Afghanistan's context?
2. How to implement EDM in Afghanistan education system effectively?
3. What is the status with regards to availability of data in Afghanistan education systems?

This paper, initially presents a description on big data, data mining, and educational data mining. Next, it briefly explains the education systems and the existence and availability of large amounts of data in Afghanistan education systems. Finally, it presents studies in the area of EDM and discusses their applicability in Afghanistan's education environment using case studies as a support for the argumentation.

## 1.1 Big Data

Upon entering the 21$^{st}$ century, there has been a massive increase in data generation with the prevalence of mobile technologies, social media networks, sensor devices for environmental data gathering, and online shopping sites, to name a few. These vast amounts of generated data are too big and complex to be handled and processed effectively by traditional approaches [34, 35]

Big data does not just mean very large amounts of data, though that could be one of the factors and characteristics of it. Big data is identified by the following three main cornerstones known as 3 V's: Volume, Variety and Velocity [34, 35]. Recently other V's were added to define the term big data very precisely such as Value, Veracity, and Visualization [32, 16].

In a nutshell, big data refers to rapidly growing structured and unstructured data with sizes beyond the ability of traditional database tools (RDBMSs) to store, manage, and analyze. Therefore, it leads to new big data solution technologies. For example, NoSQL, Apache Hadoop, Apache Spark and Apache Flink framework with ecosystems on top of each to effectively store and process the big data either in batch or streaming processes with distributed computing in a reliable and scalable fashion [5, 8, 12, 31].

It is worth mentioning that there are other non-distributed and non-scalable software and tools available for data mining purposes, e.g., Weka, SPSS Clementine, packages for R and Python programming languages. Furthermore, it should be considered that big data solutions certainly do not replace the traditional approaches. Likewise, data mining is not only for big data.

### 1.1.1 The Three V's of Big Data

The **Volume** means very large amounts of data. For example, in 2012, approximately 2.5 Exabyte of data were generated per day [17].

**Velocity** means how rapidly data is generated. Per the InternetLiveStats every second more than 10,000 Tweets are sent, 100,000 YouTube videos are viewed, 49,000 Google searches are performed, 2,300,000 emails are sent, and 27,700 Gigabytes of Internet traffic are generated [11, 20].

**Variety** refers to the 80% total existing unstructured data e.g., videos, audios, books, news articles, blogs, and click streams data [29].

## 1.2 Data Mining

Data mining is the cross-road of Statistics, Machine Learning and Artificial Intelligence algorithms, Database and Data Warehouses [10, 25].

Data mining enables organizations to explore the past by means of univariate and bivariate exploration techniques and predict the future by means of classification and clustering techniques and methods [26].

Per the KDnuggets 2014 poll [13], data mining and analytics are applied in a wide range of industries, e.g., Banking, Healthcare, Fraud Detection, Science, Advertising, Education, and many other sectors.

## 1.3 Educational Data Mining

Researchers apply and test data mining techniques to explore and mine the data which are generated from educational settings (e.g., data from schools, universities, traces that students leave when they interact with learning management systems, and intelligent tutoring systems) to better understand students and the settings which they learn in and call it EDM [4, 6, 24].

The EDM research community has undergone remarkable growth. Since 2008 every year international conference on EDM is being held to bring together researchers from computer science, education, psychology, and statistics to analyze large datasets to answer educational research questions. In 2009 the Journal of Educational Data Mining (JEDM) was established for sharing and broadcasting research results.

The main goals of EDM is to predict students' future learning behavior, to study the effects of education support that can be realized through tutoring and learning systems, to find out new models or improve the existing ones, and finally to advance scientific knowledge about learning [4].

The Application of EDM is very broad, ranging from analysis and visualization of data, providing feedback for supporting instructors, recommendations for students, predicting student performance, detecting undesirable student behaviors, grouping students, constructing courseware, planning and scheduling, and others [23].

## 2. STATUS OF AFGHANISTAN EDUCATION SYSTEMS

General education in Afghanistan comprises K-12 (primary, secondary and high school), Islamic studies, Teacher Training, Technical and Vocational schools and institutes which is administered by the Ministry of Education (MoE). The Ministry of Higher Education (MoHE) supervises the universities which provide Bachelor, Master, and Ph.D. degree programs.

The establishment of the new democracy in Afghanistan in 2002 brought hope to people to be more optimistic towards the future. The education systems have been going through a nationwide rebuilding process, and despite obstacles, numerous public and private educational institutions were established across the country [2, 27]. The result is a substantial increase in the student enrollment rate, as reflected in (see Figure 1) [3, 28].

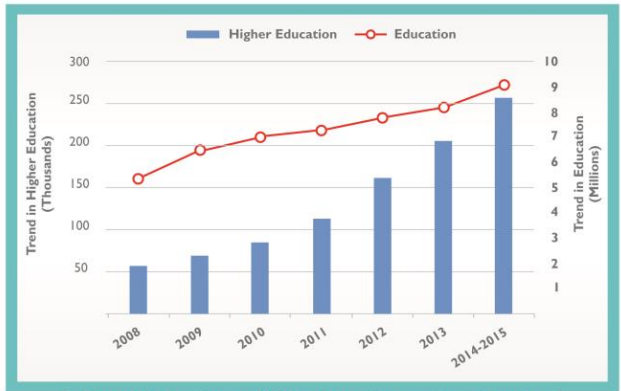

Figure 1. Education and Higher Education enrollment trends.

In 2015 around 10 million students attended more than 15,479 public and private schools [28, 30], and more than 250,000 students attended 126 public and private universities. On the other hand, every year more than 200,000 students graduate from high schools, and around 300,000 participate in the Kankor exam across the country [18, 28].

The MoE and MoHE as the main bodies of education systems in Afghanistan have been trying to standardize the quality of education to be able to meet the minimum international standards. In this extremely challenging process, one of the efforts of the MoE and MoHE has been to automate their information through Education Management Information System (EMIS) [27] and Higher Education Management Information System (HEMIS) [21].

The EMIS and HEMIS are able to generate only facts and figures which are not very helpful in decision making to improve the education systems effectively. For example, '10 million students in schools' is just a number and piece of data without a specific context and without further useful information to describe the setting.

Data needs context to be understood in its meaning and its consequences, and when data is simple, most of the times actions are impossible because they do not provide decision makers with the context in which the data was recorded. Therefore, these simple facts and figures do not help policy makers to improve the educational settings. For example, to predict proper major fields of study for high school graduates participating in the Kankor exam, to identify first year university students who are at high attrition risk, to recommend right courses for the students to enroll, and finally, to advance scientific knowledge about learners.

## 3. EDM CHALLENGES IN AFGHANISTAN EDUCATION SYSTEMS

Data is the core asset of every organization. Without presence of data, mining and analytics cannot be done, thus, organizations cannot improve their settings efficiently. It is also true that data without analysis in mind is just a bunch of data and not helpful for decision-makings.

In the context of Afghanistan education, there are around 10 million students attending schools. For each student, demographic information, economic, social and cultural status of family and kin, learning activities outcome, and performance, plus other information is stored in scattered paper-based and traditional systems. Unfortunately, most of the data are not used by the educational institutions.

Consider this scenario: students must study 12 grades to be a high school graduate. In school, on average, 13 subjects relevant to sciences, social sciences, literatures, languages, and other fields are taught per grade, and respectively, each subject comprises at least 2 exams. The following simple calculation resulted in more than 3,120,000,000 records refer (only) to the students' enrollment entity and scoring data. Of course, there are different entities and data that exist. For example, data for teachers, students and students' dependents, subjects, activities, classes, etc. are stored and the number of records will be increased dramatically.

On the other hand, to estimate the size of the data that can exist in the Afghanistan education systems, if 2 Megabytes of data in average is counted per student 20 Terabytes of data will be available for storage.

It is worth mentioning that the 20 Terabytes of data are generated from school students are explicitly stored. Imagine how much data are generated implicitly and explicitly per semester or in a year when integration of Information Technology (Learning Management Systems, Tutoring Systems or MOOCs) are introduced at schools and universities as a tool to enrich and support the education systems. The amount of data will grow intensely even exponentially.

Unfortunately, the analytics and mining processes are not used over these data to extract valuable information for better decision-making purposes to improve the education systems effectively and efficiently.

Major challenges to EDM in Afghanistan are lack of data availability and accessibility in electronic format to generate and build datasets for the mining and analytic purposes. Besides, the concept of EDM is new and research has not been done in Afghanistan. Therefore, lack of experts is another major issue.

The next section presents some of the EDM research that has been successfully applied in other countries in education systems to improve and support the educational settings. Meanwhile, the outcome of these studies and research will be highlighted in the context of Afghanistan education systems.

## 4. EDM OPPORTUNITIES IN AFGHANISTAN EDUCATION SYSTEMS

Education is the main pillar of every society. Per Hamid Karzai (the former president of Afghanistan) and Asadullah Hanif (the minister of education), education is the only way to develop and build Afghanistan. Thus, the application of EDM is a great asset for Afghanistan education systems to improve the education settings and increase the quality of education.

This section presents studies in the area of EDM and discusses the applicability, contextualization and modification in Afghanistan Education environment.

### 4.1 Case Study I: Major Prediction and Recommendation

One of the studies [22] aimed to predict student placement in class using data mining in Indonesia. Per the Indonesian government regulations, senior high school students must be divided into Science and Social Science majors, which are called Placement Class process. Traditionally, this division process is conducted by teachers. It is a difficult and time-consuming task to identify and divide the students into right and proper majors.

The researcher's main aim was to find a very promising solution to facilitate the Placement Class process. Knowledge Discovery in Databases (KDD) or data mining in education classification techniques were introduced and used to divide the students into appropriate majors.

The outcome of the research after testing different methods was automation and prediction of the Placement Class process with a 79.16% accuracy rate.

## 4.2 Case Study I: Lessons-Learned and Applicability in Afghanistan

Presently in Afghanistan, school students are not divided into Majors. The author conducted one online survey to public and private university students and graduates, and another survey to computer science students and graduates. A total of 333 people participated in these surveys; 315 agreed that it is more useful if the students are offered specialized studies after grade 9 at school. Additionally, due to general studies and insufficient orientation on Kankor at schools, the majority of students do not know what Major to choose in the Kankor. This was confirmed by the same online surveys. Besides, in the existing situation, it is found that there are no structural and specialized institutions to provide and guide students on career choices based on their skills and interests. This situation creates a vicious cycle for misappropriating human-capital as the most vital resource for development.

The outcome of this research [22] can be customized and used to recommend proper Majors to high school graduates prior attending the Kankor, and also while specialized studies are introduced at schools. The following approaches can be used. 1-Assess student performance for 10th, 11th and 12th grades to identify the strengths and weaknesses of the applicants in all the relevant Majors. 2-Since the results of high school grades could be misleading, this research also proposes the design of a new standardized test to evaluate the interest and capabilities of the applicants through varied 'Yes' and 'No' intelligent questions. 3-Since there are no pre-collegiate courses prior to entering University, it is deemed efficient to evaluate the skills of applicants in the 12th grade through several Kankor practice tests. 4-Other simulator (self-assessment) tools as an all-encompassing medium to self-evaluate, capitalize on improving and minimize the identified gaps of candidates and to evaluate the interest and capabilities of the applicants. 5-Of course, social, economic, and literacy status of student's family and other pedagogical factors could be significant for better evaluation and assessment. 6-Divide more than 100 Majors into main major areas including Natural and Social Sciences, Health Sciences, Humanities and Literature, Islamic Education, Fine Arts and Technical Education. 7-Finally, consideration of previous Kankor results data during data mining process would lead to better accuracy rate. The primary and basic diagram to predict field of study is illustrated (see Figure 2).

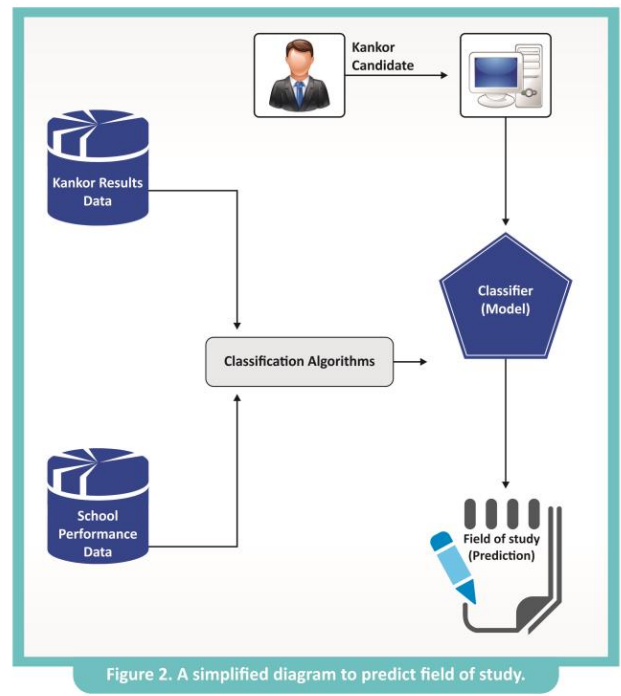

Figure 2. A simplified diagram to predict field of study.

## 4.3 Case Study II: Support at Risk Students

Another research [1] has been conducted using EDM methods and applied at the New York Institute of Technology (NYIT) to design a first-year at-risk model for NYIT to identify 1-the key factors that place a student at risk of attrition, and 2-the failure rate for each new fresh student for timely intervention before it is too late and to effectively support them accordingly. This model was built using previous historical data and went into production at NYIT with an accuracy rate of 75%.

Moreover, there has been additional research and studies [7, 15, and 19] discovered that timely intervention with students at highest risk of attrition could be effective in improving retention rates.

## 4.4 Case Study II: Lessons-Learned and Applicability in Afghanistan

There has been significant increase in enrollment rate in higher education institutions, but most of the students are at risk of dropping out with poor performance during their higher education studies. One of the main reasons is that the participants randomly select fields of study in the Kankor exam without much knowledge of the requirements and challenges ahead of them and the inventory of their existing knowledge in the relevant field of study. Additionally, lack of specialized studies at schools is another major reason for attrition and poor performance in higher education. For example, per the previous online survey conducted by the author among Computer Science students in Herat province out of 227 respondents (e.g., freshmen, sophomore, and graduates of male and female students) around 90% did not have the skill and knowledge of programming, database concept, and operating systems, as echoed in (see Figure 3).

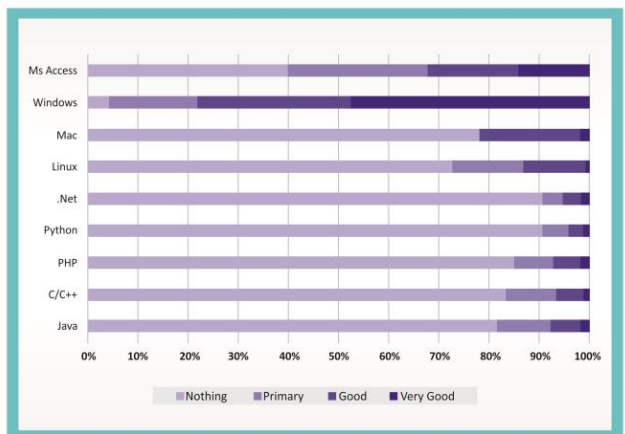

Figure 3. IT skill of computer science students prior Kankor.

The result of the survey is showing that one of the major reasons for weak academic performance in higher education is lack offering specialized studies in school.

An early counseling intervention solution would be a great support to identify the key factors to improve their academic performance and to decrease rates of attrition through academic counseling, tutorial classes and other supportive programs [1]. This could be achieved with evaluation and comparison of fresh student's data with historical data of senior students. For example, school performance and grades for main prerequisite subjects relevant to their selected Major (i.e. the required score value for Journalism in mathematics might be 2 out of 5, while in Engineering it might be, 5 out of 5), if they attended supportive courses and classes besides school studies, family responsibilities, and other social and extracurricular activities.

## 4.5 Case Study III – Recommender Systems in Education

The recommender systems have evolved in the extremely interactive environments of the web, particularly, e-commerce and other popular and social platforms. They apply data mining and analyses techniques to the problem of helping customers find which products they would like to purchase at e-commerce sites. For example, a recommender system on Amazon.com (http://www.amazon.com) suggests other similar products to the customers that they might wish to purchase based on their previous purchasing history and other customers who purchased similar products. Until 2002, learning management systems lacked a recommendation feature to assist enrolled students to enhance online-learning processes [33]. Therefore, researchers studied the application and the potential of recommendation in the education domain, and specially in Massive Open Online

Courses (MOOCs) which are a relatively new phenomenal widespread higher education.

Most of today's learning management systems are very popular and store vast logs of students' performance and behavior implicitly or explicitly while students are browsing courses, taking tests and quizzes, reading news, watching videos, following forum threads and topics, etc. This data includes the session duration of entry and exit time, mouse movements, delay taking the tests and quizzes, visiting pages and modules, etc. Using data mining techniques on these collected data can recommend students proper and related courses. It also enables the system to predict the students' score and result while choosing courses before even any examination processes [14, 33].

## 4.6 Case Study III: Lessons-Learned and Applicability in Afghanistan

Over the past decade, Afghanistan has experienced an extremely rapid growth in ICT infrastructure and services. More than 85% of the population (17.4 million subscribers) geographically has GSM coverage and 3G services. In 2012, the total internet users were estimated to be 1 million [9]. Therefore, the development within the ICT infrastructures as well as availability of user-friendly open source learning platforms i.e. Moodle (https://www.moodle.org) has made the environment ready to gradually transfer paper-based services to online-based services, particularly, within the higher education institutions to enrich and support the current education system with integration of Information Technology. This approach enables the educational institutions to collect the data implicitly and explicitly for the mining and improvement purposes.

## 5. CONCLUSION

Enrolment trends in Education and Higher Education generates vast amounts of data. With learning and tutoring management systems, the amount of data will be significantly increased either implicitly or explicitly. The main challenge preventing the applicability of EDM in Afghanistan is lack of proper data storage and accessibility to data in electronic format. EMIS at MoE and HEMIS at MoHE together could be appointed to provide the raw data for EDM applications to help discern patterns of abilities and behaviors which could be used to help educational institutions.

## 6. OUTLOOK

There are discussions implementing and applying distance and e-Learning in Afghanistan higher education. Prior to introducing E-Learning and offering fully online courses; environment readiness is important. The environment readiness can be achieved by introducing learning and tutoring platforms as part of the education and higher education institutions to enrich and support the current education system. Moreover, the application of data mining will improve the traditional education systems. The former factor brings society awareness and the later one enlightens and strengthens academicians mining the data and the importance of data availability and accessibility. Then after proper infrastructure is in place, introducing e-Learning and offering blended or fully online courses would be more beneficial.

MoE is seeking support to integrate all the schools across the country under one union online platform in its Center of Science campus. This is a great opportunity to think of an in-depth mechanism to collect students' individual information both for transactional processing and analytical and mining processing.

Finally, since education is foundation for Higher education. Therefore, strong and close cooperation between MoE and MoHE is crucial to increase the quality of Higher education. MoE should have a mechanism to share the archived data with MoHE during the Kankor exam as well as when students get admission to universities for mining purposes to support students more effectively and efficiently.

## 7. ACKNOWLEDGMENTS

I thank my supervisors at Technical University of Berlin (Prof. Dr.-Ing. Uwe Neumann, Prof. Dr. Sebastian Bab, and Dr. Nazir Peroz) for their direct and indirect support, and the respondents.

## 8. REFERENCES

[1] Agnihotri Lalitha, Ott Alexander. 2012. Building a Student At-Risk Model: An End-to-End Perspective. In Proceedings of the 7th International Conference on Educational Data Mining, 209-212

[2] Andishman Mohammad Ikram. 2010. Modern Education in Afghanistan. Maiwand publication

[3] Aturupane Harsha. 2013. Higher Education in Afghanistan: an emerging mountainscape. Retrieved May 25, 2015 from http://documents.worldbank.org/curated/en/2013/08/18197239/higher-education-afghanistan-higher-education-afghanistan-emerging-mountainscape

[4] Baker S.J.D. Ryan, Yacef Kalina. 2009. The State of Educational Data Mining in 2009: A Review and Future Visions. Journal of Educational Data Mining 1, 1:3-16

[5] Dean Jeffrey, Ghemawat Sanjay. 2004. MapReduce: simplified data processing on large clusters. In Proceeding OSDI'04 Proceedings of the 6th conference on Symposium on Operating Systems Design & Implementation, 10-10

[6] EDM. 2015. Educational Data Mining. Retrieved May 15, 2015 from http://www.educationaldatamining.org

[7] Fowler R. Paul, Boylan R. Hunter. 2010. Increasing Student Success and Retention: A Multidimensional Approach. Journal of Developmental Education 34, 2:2-4, 6, 8-10

[8] Ghemawat Sanjay, Gobioff Howard, Leung Shun-Tak. 2003. The Google file system. In Proceeding SOSP '03 Proceedings of the nineteenth ACM symposium on Operating systems principles, 29-43.

[9] Hamdard Jawid. 2012. The State of Telecommunications and Internet in Afghanistan - Six Years Later (2006-2012). Retrieved May 10, 2015 from https://www.internews.org/sites/default/files/resources/Internews_TelecomInternet_Afghanistan_2012-04.pdf

[10] Han Jiawei, Kamber Micheline, Pei Jian. 2011. Data Mining Concepts and Techniques. Morgan Kaufmann

[11] Internet Live Stats. 2015. One Second. Retrieved May 21, 2015 from http://www.internetlivestats.com/one-second/

[12] Karau Holden, Konwinski Andy, Wendell Patrick, Zaharia Matei .2015. Learning Spark: Lightning-Fast Big Data Analysis. O'Reilly Media

[13] Kdnuggets. 2014. Industries Applied Analytics Data Mining and Data Science. Retrieved June 02, 2015 from http://www.kdnuggets.com/polls/2014/industries-applied-analytics-data-mining-data-science.html

[14] Maghsoudi Behroz, Soleimani Sadeq, Amiri Ali, Afsharchy Mohsen. 2012. Improving education quality in e-learning systems using data mining. Scientific Information Database 6, 4:276-278

[15] MÁRQUEZ-VERA C., ROMERO C., VENTURA S. 2011. Predicting School Failure Using Data Mining. Journal of Educational Data Mining

[16] Marz Nathan, Warren James. 2015. Big Data: Principles and best practices of scalable realtime data systems. Manning Publications

[17] McAfee Andrew, Brynjolfsson Erik. 2012. Big Data: The Management Revolution. Retrieved May 15, 2015 from https://hbr.org/2012/10/big-data-the-management-revolution/ar

[18] Ministry of Higher Education. 2014. Universities. Retrieved May 29, 2015 from http://mohe.gov.af/university/en

[19] Pan Wei, Guo Shuqin, Alikonis Caroline, Bai Haiyan. 2008. Do Intervention Programs Assist Students to Succeed in College?: A Multilevel Longitudinal Study. College Student Journal 42, 1: 90-98

[20] Pennystocks. 2015. Internet in Real Time. Retrieved May 21, 2015 from http://pennystocks.la/internet-in-real-time/

[21] Peroz Nazir, Tippmann Daniel. 2012. Information Technology for Higher Education in Afghanistan: ZiiK Report Nr. 32. Retrieved May 25, 2015 from https://www.ziik.tu-berlin.de/menue/veroeffentlichungen/reports/

[22] Pratiwi O.N. 2013. Predicting student placement class using data mining. In Teaching, Assessment and Learning for Engineering (TALE), 2013 IEEE International Conference on, 618-621, 26-29

[23] Romero C., Ventura S. 2010. Educational Data Mining: A Review of the State of the Art. In Systems, Man, and Cybernetics, Part C: Applications and Reviews, IEEE Transactions on, 601-618

[24] Romero Cristobal, Ventura Sebastian, Pechenizkiy Mykola, Baker S.J.d. Ryan. 2010. Handbook of Educational Data Mining. CRC Press

[25] Sayad Saed. 2011. Real Time Data Mining. Self-Help Publishers

[26] Sayad Saed. 2015. An Introduction to Data Mining. Retrieved May 15, 2015 from http://www.saedsayad.com

[27] Silva DE. Samantha. 2015. The Impact of Education Management Information Systems: The Case of Afghanistan. Retrieved May 29, 2015 from http://blogs.worldbank.org/education/impact-education-management-information-systems-case-afghanistan

[28] The Central Statistics Organization. 2014-2015. Afghanistan Statistical Yearbook 2014-2015: Education Part One. Retrieved June 15, 2015 from http://cso.gov.af/en/page/1500/4722/2014-2015

[29] The CRISIL Global Research & Analytics. 2013. Big Data - The Next Big Thing. Retrieved September 17, 2015 from http://www.crisil.com/global-offshoring/gra-nasscom.html

[30] The Ministry of Education. 2014. Report. Retrieved May 29, 2015 from http://moe.gov.af/en/page/1831/3031

[31] White Tom. 2015. Hadoop: The Definitive Guide. O'Reilly Media

[32] Wolff J.G. 2014. Big Data and the SP Theory of Intelligence. Access, IEEE 10, 2:301-315

[33] Zaiane O.R. 2002. Building a recommender agent for e-learning systems. In proceedings of International Conference on Computers in Education, 55-59

[34] Zibin Zheng, Jieming Zhu, Lyu R. Michael. 2013. Service-Generated Big Data and Big Data-as-a-Service: An Overview. IEEE International Congress: 403-410

[35] Zikopoulos Paul, deRoos Dirk, Bienko Christopher, Andrews Marc, Buglio Rick. 2014. Big Data beyond the Hype: A Guide to Conversations for Today's Data Center. McGraw-Hill Professional